\documentclass{basi}
\usepackage{txfonts}
\usepackage{graphicx}
\def\be{\begin{equation}}
\def\ee{\end{equation}}
\def\beq{\begin{eqnarray}}
\def\eeq{\end{eqnarray}}
\newcommand{\beqn}{\begin{eqnarray*}}
\newcommand{\eeqn}{\end{eqnarray*}}
\def\ben{\begin{enumerate}}
\def\een{\end{enumerate}}
\newcommand{\dps}{\displaystyle}
\input dpscolor.sty
\def\op{ \ $ }
\def\cl{$ \ }
\def\evec{\vec{\hbox{\bf E}}}
\def\sch{Schwarzschild }

\def\msun{M_\odot}

\begin{document}
\title[Stellar perturbations]{Gravitational waves from perturbed stars}
\author[V.~Ferrari]
       {V.~Ferrari\thanks{e-mail:valeria.ferrari@roma1.infn.it}\\
Dipartimento di Fisica G. Marconi, Sapienza Universit\`a di Roma\\
and Sezione INFN ROMA1, Piazzale Aldo Moro 5, 00185 Roma, Italy\\
       }

\pubyear{2011}
\volume{39}
\pagerange{\pageref{firstpage}--\pageref{lastpage}}
\date{Received 2011 February 11; accepted 2011 February 28}

\maketitle
\label{firstpage}

\begin{abstract}
Non radial oscillations of neutron stars are associated with the emission of 
gravitational waves. The characteristic frequencies of these oscillations
can be computed using the theory of stellar perturbations, and they are shown to
carry
detailed information on the internal structure of the emitting source.
Moreover, they appear to be encoded in various radiative processes, as for instance 
in the tail of the giant flares of Soft Gamma Repeaters. Thus, their determination is 
central to the theory of stellar perturbation. A viable  approach  to the
problem consists in formulating this theory as a problem of resonant
scattering of gravitational waves incident on the potential barrier
generated by the spacetime curvature. This approach discloses some
unexpected correspondences between the theory of stellar perturbations and
the theory of quantum mechanics, and allows us to predict new relativistic effects.

\end{abstract}


\section{Introduction}\label{s:intro}
The theory of stellar perturbations is a very powerful tool to investigate the
features of gravitational signals emitted  when a star is set in non-radial
oscillations by any external or internal cause. The
characteristic frequencies at which waves are emitted are of great interest in
these days, since gravitational wave detectors Virgo and LIGO
are approaching the sensitivity
needed to detect gravitational waves emitted by pulsating stars.  These
frequencies carry information on the internal structure of a star, and 
appear to be encoded in various radiative processes; thus,
their study is a central problem in perturbation theory and in
astrophysics.
In this paper I will  illustrate  the theory of perturbations of a 
non-rotating star, describing in particular the formulation  that Chandrasekhar 
and I developed, its motivation and outcomes. 
Furthermore, I will briefly describe how the theory has been applied to study the
oscillation frequencies of neutron stars.

In order to frame the problem in an appropriate  historical perspective,
it is instructive to remind ourselves how the study of stellar perturbations
was treated in the framework of Newtonian gravity.
In that case,  the adiabatic
perturbations of a spherical star are described by a fourth-order, linear,
differential system which couples the perturbation of the Newtonian
potential to those of the stellar fluid.
All perturbed quantities are Fourier-expanded and, after a suitable expansion in
spherical harmonics, which allows for the separation of variables, the relevant
equations are manipulated in such a way that the quantity which is singled out to
describe the perturbed star is the Lagrangian displacement $\vec{\xi}$ experienced by
a generic fluid element; indeed, the changes in 
density, pressure and gravitational potential induced by the perturbation, 
can all be expressed uniquely in terms of $\vec{\xi}$.
The equations for  $\vec{\xi}$ have to be solved by imposing appropriate boundary
conditions at the centre of the star, where  all physical quantities must be regular,
and on its boundary, where the perturbation of the pressure must vanish.  These
conditions are satisfied only for a discrete set of {\it real} values of the
frequency, $\{ \omega_n\}$,  which are the frequencies of the star's {\it normal modes}.
Thus, the linearized version of the Poisson and of the hydro equations are reduced to
a characteristic value problem for the frequency $\omega$. 

An adequate base for a rigorous  treatment of stellar pulsations of a spherical star
in general relativity was provided by K.S. Thorne and collaborators in  a series of
papers published in the late sixties--early seventies of the last century
(Thorne \& Campolattaro 1967, 1968; Campolattaro \& Thorne 1970; Thorne 1969a,b; 
Ipser \& Thorne 1973). The theory was developed in
analogy with the Newtonian approach, and was later completed  by Lindblom \&
Detweiler (1983), who brought the analytic framework
to a form suitable for the numerical integration of the equations, thus allowing for
the determination of the real and imaginary part of the characteristic frequencies of
the $\ell=2$, {\it quasi-normal modes}.  Indeed, a main difference between the
Newtonian and the relativistic theory is that in general relativity the oscillations
are damped by the emission of gravitational waves, and consequently the mode
eigenfrequencies are complex.  Higher order ($\ell >2$) mode frequencies were
subsequently computed by Cutler \& Lindblom (1987).

In 1990, Professor Chandrasekhar and I started to work on stellar perturbations, and
we decided to derive ab initio the equations of stellar perturbations following a
different approach, having as a guide the theory of black hole perturbations rather
than the Newtonian theory of stellar perturbations. 

\subsection{Black hole perturbations: wave equations and conservation laws}
\begin{figure}[ht]
\begin{center}
\includegraphics[angle=270,width=.6\textwidth]{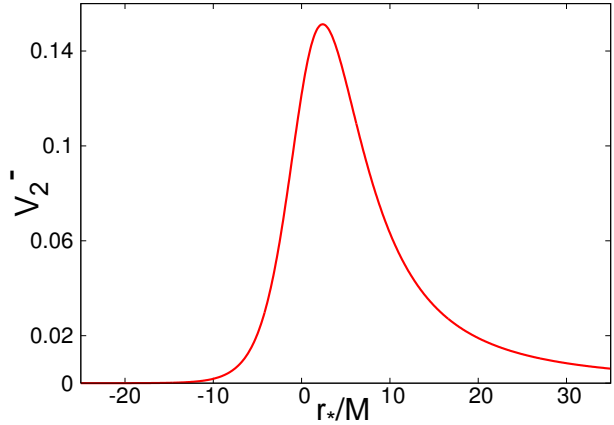}
\end{center}
\caption[]{The potential barrier generated by the axial perturbations of a
Schwarzschild black hole for $\ell =2$. The potential for the polar perturbations 
has a similar form.}
\label{fig1}
\end{figure}

In 1957 T. Regge and J.A. Wheeler 
set the basis of the theory black hole perturbations
showing that, by expanding the metric perturbations of a Schwarzschild black hole
in tensorial spherical harmonics, Einstein's 
equations can be separated (Regge \& Wheeler 1957).
Spherical harmonics belong to two different classes, depending on the way they
transform under the parity transformation $ \theta\rightarrow\pi-\theta$ and $
\varphi\rightarrow\pi+\varphi$; those which transform like $ (-1)^{(\ell+1)}$ are
named {\it odd}, or {\it axial},  those that transform like $ (-1)^{\ell}$ are
named {\it even}, or {\it polar}.  The perturbed equations decouple in
two distinct sets belonging to the two parities.
Regge \& Wheeler further showed that by
Fourier-transforming the time dependent variables,
the equations describing the radial part of the {\it axial} perturbations can 
easily be reduced  to a single Schroedinger-like equation, and 13 years later 
Frank Zerilli showed that this can also be done for the much more 
complicated set of {\it polar} equations (Zerilli 1970a,b); thus, the axial and polar
perturbations of a Schwarzschild black hole are described by the wave equation 
\beq
\label{reggew}
&&\frac{d^2 Z_\ell^\pm}{dr_*^2}+
\left[\omega^2-V^\pm_\ell(r)\right]Z_\ell^\pm=0~,
\\\label{potrw}
&&V^-_\ell(r)=\frac{1}{r^3}\left(1-\frac{2M}{ r}\right)[\ell(\ell+1)r-6M]
\\
&&V^+_\ell(r)=
\frac{2(r-2M)}{r^{4}(nr+3M)^2}
[n^2(n+1)r^3+3M n^2 r^2+ 9M^2nr+9M^3]~.
\label{potzer}
\eeq
where $r_*=r+2M\log\left(\dps\frac{r}{2M}-1\right)$,
$n=\frac{1}{2}(\ell+1)(\ell-2)$, and
$M$ is the black hole mass. The superscript $-$ and $+$ indicate, respectively,
the Regge-Wheeler equation for the axial perturbations, and the Zerilli equation for
the polar perturbations, and the corresponding potentials.
The Regge-Wheeler potential for $\ell=2$ is shown in figure \ref{fig1}.

An alternative approach to studying black hole  perturbations 
considers   the perturbations of the Weyl and Maxwell scalars within
the Newman-Penrose formalism.
Using this approach in 1972  S. Teukolsky 
was able to  decouple and separate the equations governing 
the perturbations of a Kerr black hole (Teukolsky 1972, 1973), 
and to reduce them to a single master equation
for the radial part of the perturbation  $ R_{lm}$:
\be
\label{teuk}
\Delta  R_{lm,rr}+2(s+1)(r-M)  R_{lm,r}+
V(\omega,r) R_{lm}=0~,\qquad
\Delta =r^2-2Mr+a^2.
\ee
Variable separation was achieved in terms of oblate spheroidal harmonics, and 
the potential $V(\omega,r)$ is given by
\beq
&&V(\omega,r)=
\frac{1}{\Delta}\left[ (r^2+a^2)^2\omega^2-4aMrm\omega
+a^2m^2+
2is(am(r-M)-M\omega(r^2-a^2))
\right]\\\noindent
&&+\left[ 2is\omega r-a^2\omega^2-A_{lm}\right]~,
\label{pott}
\eeq
where $a$ is the black hole angular momentum, $A_{lm}$ is a separation constant, and $s$, the  spin-weight parameter,
takes the values $s=0,\pm 1,\pm 2,$ respectively  for scalar, electromagnetic and
gravitational perturbations.
It may be noted that for the Schwarzschild perturbations, due to the 
background spherical symmetry,
non-axisymmetric modes with a $e^{im\phi}$ dependence
can be deduced from axisymmetric, $m=0$ modes by suitable rotations of the polar axes.
As a consequence, the potentials (\ref{potrw}) and (\ref{potzer}) do not depend on 
the harmonic index $m$.
Conversely, since Kerr's background is axisymmetric,
this degeneracy is removed and the potential (\ref{pott}) depends on $m$.
Moreover, while the Schwarzschild  potentials (\ref{potrw}) and (\ref{potzer})
are  real and independent of frequency,
the potential barrier of a Kerr black hole is complex, and depends on
frequency.

The wave equations governing black hole perturbations show that the curvature
generated by a black hole appears in the perturbed equations
as a one-dimensional potential barrier;
consequently, the response of a black hole to a generic perturbation
can be studied by investigating the manner in which a gravitational wave incident
on that barrier  is transmitted, absorbed and reflected.
Thus, the theory of black hole perturbations can be formulated as a scattering
theory, and the methods traditionally applied in quantum mechanics to investigate 
the behaviour of physical systems described by a Schroedinger
equation can be adapted and used to study the behaviour of perturbed black holes. 
For instance, it is known that in quantum mechanics, given
a one-dimensional potential barrier associated with a Schroedinger equation,
the singularities in the scattering cross-section
correspond to complex eigenvalues of the energy and to the so-called quasi-stationary
states. Since the perturbations of a Schwarzschild black hole are described by the
Schroedinger-like equation (\ref{reggew}) with  the one-dimensional potential barriers
(\ref{potrw}) and  (\ref{potzer}), in which the
energy is replaced by the frequency, the singularities in the scattering
cross-section will provide the complex values of the black hole eigenfrequencies,
and the corresponding eigenstates will be the black hole Quasi Nomal Modes (QNM).
These modes  satisfy the boundary conditions of a pure
outgoing gravitational wave emerging at radial infinity $(r_* \rightarrow +\infty)$,
and a pure ingoing wave 
impinging at the black hole horizon $(r_* \rightarrow -\infty)$.
That these solutions should exist had been suggested by
C.V. Vishveshwara  in 1970 (Vishveshwara  1970), and the next year
W.H. Press  confirmed this idea  by numerically
integrating the axial wave equation (\ref{reggew}), and by showing that
an arbitrary initial perturbation
ends in a ringing tail, which indicates that  black holes
possess some proper modes of vibration (Press 1971).
However, it was only in 1975 that S. Chandrasekhar and S.  Detweiler
computed the complex eigenfrequencies
of the  quasi-normal modes of a Schwarzschild black hole,
by integrating the Riccati equation associated with
the axial equation (\ref{reggew}) (Chandrasekhar \& Detweiler 1975).
In addition, they  also
showed that the transmission and the reflection coefficients
associated respectively  with the polar and with the axial
potential barriers are equal. As a consequence, the polar and the axial 
perturbations are isospectral, i.e. the polar and axial QNM eigenfrequencies are
equal.  This equality can be explained in terms of a transformation theory
which clarifies the relations that exist between
potential barriers admitting
the same reflection and absorption coefficients.
This is an example of how the scattering approach has been effective not only 
to determine the QNM frequencies, but also to investigate the inner relations
existing among the axial and polar potential barriers and to gain a deeper insight 
in the mathematical theory of black holes, which was
illustrated by S. Chandrasekhar in his book on the subject
(Chandrasekhar 1984).
Following this approach, a variety of methods developed in the context of quantum
mechanics have been used to determine the QNM spectra of rotating and non-rotating
black holes, such as the WKB and higher
order WKB method, phase-integral methods and the theory of Regge poles, 
just to mention some of them (Schutz \& Will 1985; Ferrari \& Mashhoon 1984a,b;
Andersson, Araujo \& Schutz 1993a,b,c; Andersson 1994; Andersson \& Thylwe 1994).
\subsection{A conservation law for black hole perturbations
and its generalization to perturbed stars \label{sec_conserv}}
In quantum mechanics  the equation
\be \label{conse}
\vert  R\vert^2+\vert  T\vert^2=1,\ee
where $R$ and $T$ are the reflection and transmission coefficients associated with a
potential barrier,
expresses the symmetry and unitarity of the scattering matrix; 
it says that, if a wave of unitary amplitude is incident on one side of the potential
barrier, it gives rise to a reflected and a transmitted
wave such that the sum of the square of
their amplitudes is still one. Therefore, equation (\ref{conse}) is an energy conservation
law for the scattering problem described by the Schroedinger equation with a
potential barrier.
This conservation law is a consequence of  the constancy of
the Wronskian of pairs of independent solutions of the
Schroedinger equation. Similarly, the constancy of the Wronskian of two
independent solutions of the black hole wave equations allows us to write the same
relation between the reflection and transmission coefficients associated with the potential
barrier, and therefore it shows that such an energy conservation law also governs 
the scattering of gravitational waves by a perturbed black hole.
It should be stressed that such energy conservation law 
{\it does not} exist in the framework of the exact non-linear theory; however,
it can be derived in perturbation theory
both for Schwarzschild, Kerr and Reissner-Nordstrom  black holes.

This possibility led  Chandrasekhar to the following consideration.
Since in general relativity, any distribution of matter
(or more generally energy of any sort) induces a curvature of 
the spacetime --  a potential well --  instead
of picturing the non-radial oscillations of a star as caused by some
unspecified external perturbation, we can picture them as excited by
incident gravitational radiation.
Viewed in this manner, the reflection
and absorption of incident gravitational waves by black holes and the
non-radial oscillations of stars, become different aspects of
the same basic theory.  However, this idea needed to be substantiated 
by facts, and our starting point was to show that also for perturbed stars it is
possible to write an energy  conservation law in terms of Wronskians of independent
solutions of the perturbation equations of a spherical star.
This is easy if we consider the axial perturbations of a non-rotating star, because
in that case, as we shall show in Section \ref{secax}, the perturbed equations can be
reduced to a wave equation with a one dimensional potential barrier as for black
holes. However, to derive the conservation law for the polar perturbations was not
easy, because the corresponding equations
are  a fourth order linear differential system, in which the perturbed metric
functions couple to the fluid  perturbations, and it was not clear how to
define the conserved current. Anyway, working hard on the equations,
we were able to derive a  vector $\evec$  in terms of
metric and fluid perturbations, which satisfies the following
equation (Chandrasekhar \& Ferrari 1990a):
\be
\label{flu2}
\frac{\partial}{\partial x^\alpha}E^{\alpha}=0, \qquad
\alpha=(x^2=r,x^3=\vartheta).
\ee
(It is worth reminding ourselves that, due to the spherical symmetry, it is not restrictive to
consider axisymmetric perturbations $m=0$).
The vanishing of the ordinary divergence implies that, by Gauss's theorem, the flux
of $\evec$ across a closed surface surrounding  the star is a constant.
When the fluid variables are switched off, this conservation law reduces to that
derived for a Schwarzschild black hole, and therefore we thought 
we were on the right track. However, there was still a  question to answer:
are we entitled to say that the vector $\evec$ actually  represents the
flux of gravitational energy which develops through the stars
and propagates outside? If so, equation (\ref{flu2}) 
should reduce to the second variation of the
time component of the well known equation
\be
\label{cons2}
\frac{\partial}{\partial x^\nu}\left[\sqrt{-g}\left(
T^{\mu\nu}+t^{\mu\nu} \right) \right]=0.
\ee
where $t^{\mu\nu}$ is the stress-energy pseudotensor  of the gravitational field.
The problem is that $ t^{\mu\nu}$ is not uniquely defined; indeed
equation  (\ref{cons2}) shows that it is defined up to a divergenceless term.
A possible definition is that given by Landau \& Lifschitz (1975),
which has the advantage of being symmetric.
However, the second variation of the time component of
equation (\ref{cons2}) assuming $t^{\mu\nu}=t^{\mu\nu}_{LL}$,
does not give the divergenceless equation satisfied by our vector $\evec$,
neither for the Einstein-Maxwell case, nor in the
case of a star.
Then, Raphael Sorkin suggested  that the pseudo-tensor  whose second
variation should reproduce our conserved current is
the Einstein pseudo-tensor, because its second variation retains its divergence-free
property, provided only the equations governing the static spacetime and
its linear perturbations are satisfied\footnote{It should be mentioned 
that the first variation of the Einstein pseudo-tensor vanishes identically.}.
This property is a consequence of the  Einstein
pseudo-tensor being a Noether operator for the gravitational
field;  the Landau-Lifshitz pseudotensor failed to reproduce the
conserved current because it does not satisfy  the
foregoing requirements. In addition, Sorkin pointed
out that the contribution of the source should
be  introduced not by adding  the second variation of
the source stress-energy tensor  $T^{\mu\nu}$,
as one might  naively have thought, but through a suitably
defined Noether operator, whose form
he  derived for an electromagnetic field (Sorkin 1991).
Though this  operator does not coincide with 
$T^{\mu\nu}$, it gives the same conserved quantities.
Thus, the flux integral which we had obtained, I would say, by brute force, working
directly on the perturbed hydrodynamical equations, could be obtained from a
suitable expansion of the Einstein pseudo-tensor showing that, as for black holes,
energy conservation also governs  phenomena involving 
gravitational waves emitted by perturbed stars (Chandrasekhar \& Ferrari 1991a). 
We therefore decided to derive ab initio the equations of perturbations of a
spherical star in the same gauge used when studying the perturbations of a 
Schwarzschild black hole,
and to study the problem as a scattering problem.
In the next sections 
I shall briefly illustrate the main results we obtained by using this
approach (Chandrasekhar \& Ferrari 1990b, 1991b,c, 1992; 
Chandrasekhar, Ferrari \& Winston 1991).

\section{Perturbations of a non-rotating star \label{secax}}
\begin{figure}[ht]
\includegraphics[width=4.5cm,angle=270]{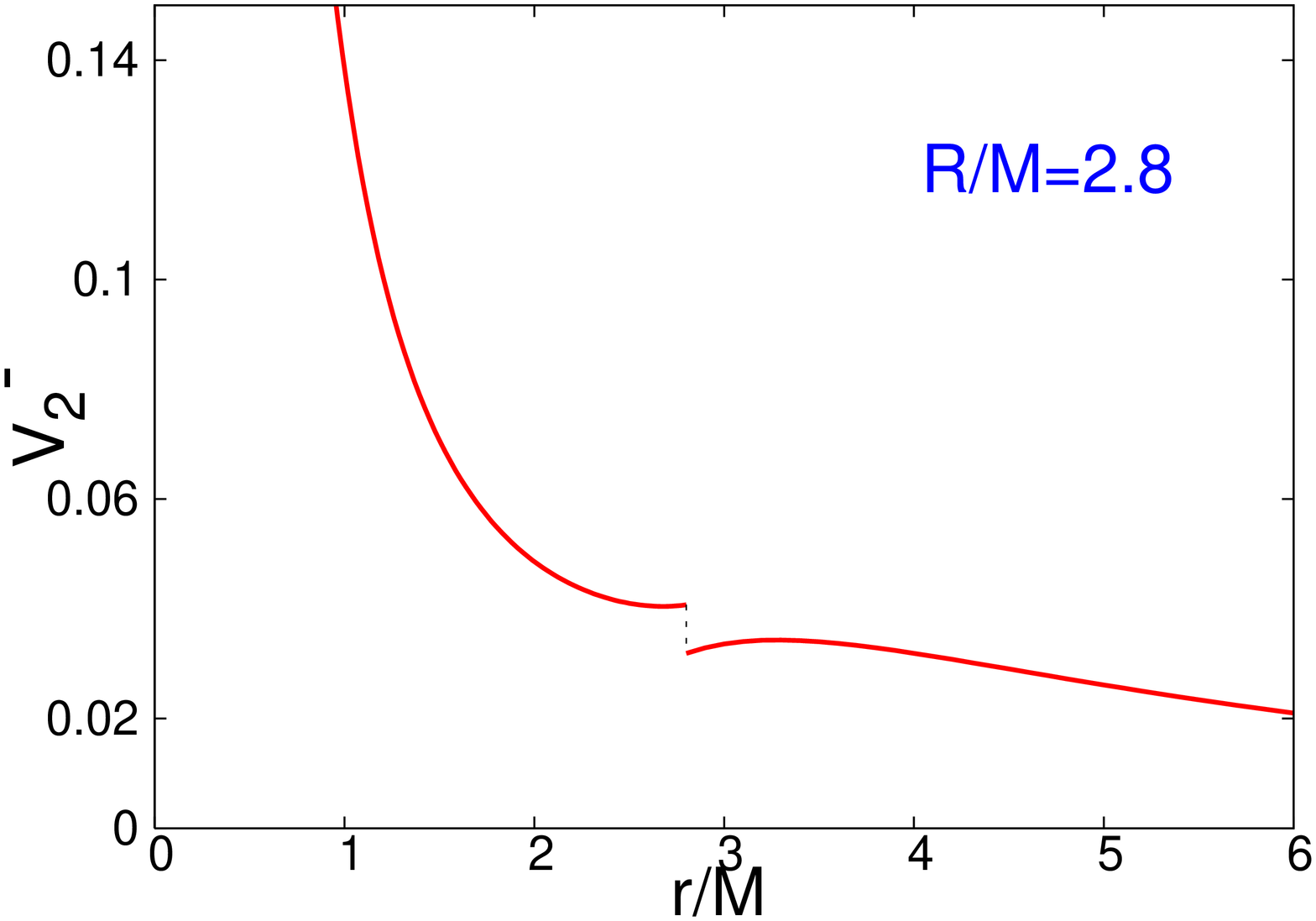}
\includegraphics[width=4.5cm,angle=270]{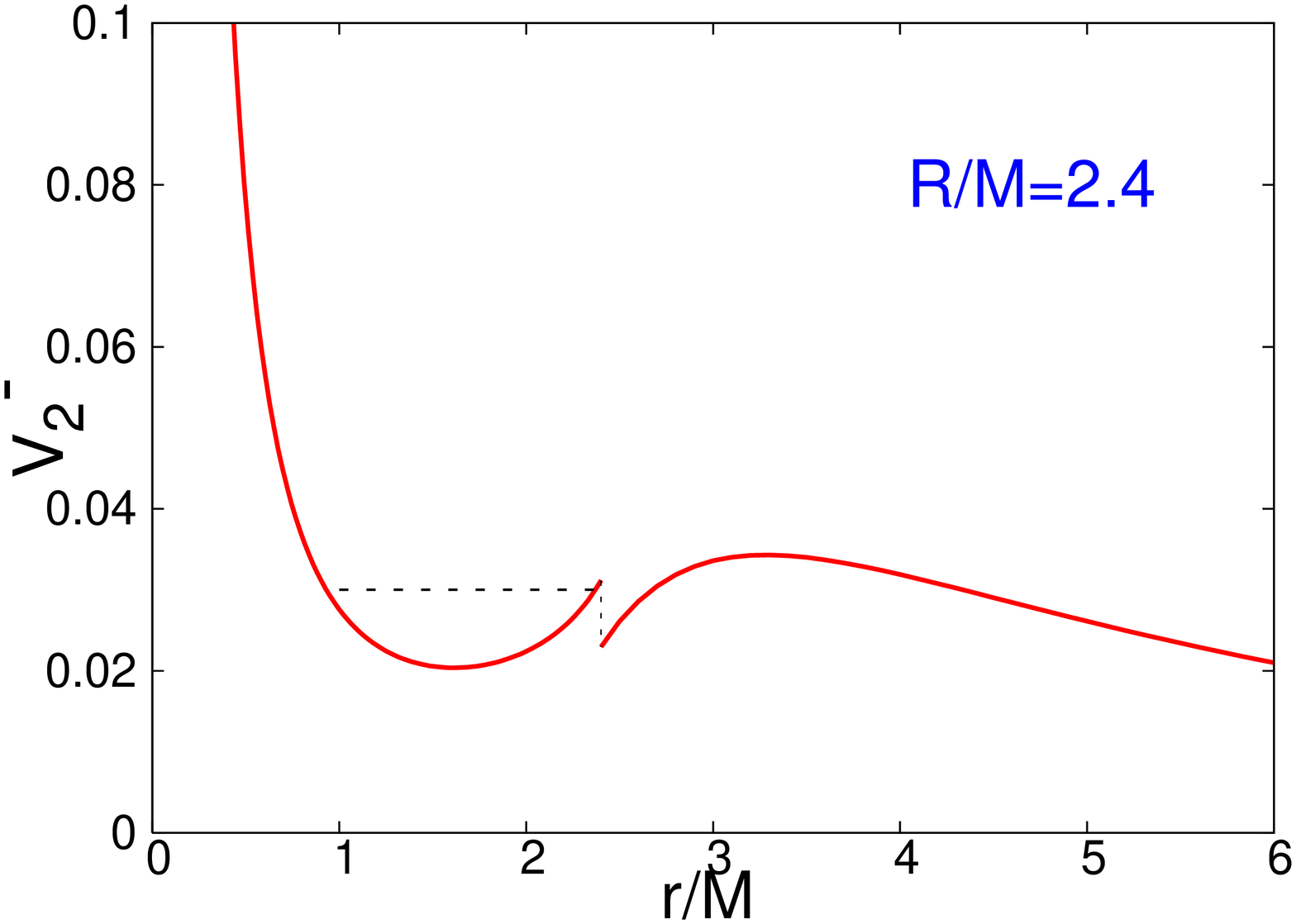}
\caption[]{The $\ell=2$  potential barrier for a constant density star of mass
$M=1.4~M_\odot$ and $R/M=2.8$ (left panel) and $R/M=2.4$ (right panel).
The horizontal, dashed line in the right panel corresponds to the value of 
frequency $\omega^2$, which corresponds to a solution regular at $r=0$ and behaving
as a pure outgoing wave at radial infinity, i.e. to a quasi-normal mode.}
\label{fig2}
\end{figure}

As for a Schwarzschild black hole, when the equations  describing the perturbations of a
spherical star are perturbed and expanded in spherical tensor harmonics they decouple in
two distinct sets, one for the polar and one for the axial perturbations.
The axial equations do not involve  fluid  motion except for a stationary rotation,
while the polar equations couple fluid and metric perturbations. The axial equations
are therefore much simpler and we showed that, after separating the variables and
Fourier-expanding the perturbed functions, they can be combined as in the
Schwarzschild case, and reduced to a single Schroedinger-like equation with a
one-dimensional potential barrier  (Chandrasekhar \& Ferrari 1990b, 1991c):
\be
\label{axial}
\frac{d^{2}Z^-_\ell}{ dr_{*}^{2}}+
[\omega^{2}-V^-_\ell(r)]Z^-_\ell=0,
\ee
where 
\be
\label{rstar}
r_*=\int_0^r e^{-\nu+\mu_2}dr,
\ee
and
\be
\label{zz1}
V^-_{\ell}(r)=\frac{e^{2\nu}}{ r^{3}}[\ell(\ell+1)r+r^{3}(\epsilon -p)-6m(r)].
\ee
The functions $\nu(r)$ and $\mu_2(r)$, which appear in the definition of the
radial variable $r_*$, are
two metric functions which are found by solving the equations of stellar structure for an
assigned equation of state (EOS).
$\epsilon(r)$ and $p(r)$ are the energy density and the pressure  in the unperturbed
star; outside the star they vanish and equation (\ref{zz1})
reduces to  the Regge-Wheeler potential (\ref{potrw}) of a
Schwarzschild black hole. Thus, the axial potential barrier generated by the 
curvature of the star depends  on how the energy-density and the pressure are distributed  
inside the star in the equilibrium configuration, and therefore it depends on the
equation of state  of matter inside the star. 
As an example, in Figure \ref{fig2} we show the $\ell=2$ potential barrier for an ideal,
constant density star with $R/M=2.8$ (left panel) and $R/M=2.4$ (right
panel).
If we compare the potential shown in Figure \ref{fig2}  with the Regge-Wheeler 
potential of a Schwarzschild black hole
 shown in Figure \ref{fig1},  we notice an important difference.  
The \sch potential vanishes at the  black
hole horizon, and has a maximum at $r_{max}\sim 3M$, whereas the potential barrier of a
perturbed star tends to infinity at $r=0$.
Thus, for a \sch black hole waves are scattered by a one-dimensional
potential barrier, whereas in the case of a star they are scattered by a
central potential.

Since the axial perturbations do not excite any motion in the fluid,
for a long time they have been  considered as trivial.  But
this is not true if we adopt the scattering approach: the absence of fluid motion
simply means that the incident axial wave experiences a potential scattering,
and this scattering can, in some extreme
conditions, be resonant.
Indeed, if we look for solutions that are regular at $r=0$ 
and behave as pure outgoing waves at infinity, we find  modes which do not 
exist in Newtonian theory;  if the star is extremely compact, the potential in
the interior is a well, and if this well is deep enough there can exist one or more 
more slowly damped quasi-normal modes, or $s$-modes (Chandrasekhar \& Ferrari 1991c).
For example, if the mass of the star is, say, $M=1.4~M_\odot$ and
$R/M=2.4$, i.e. the stellar compactness is $M/R=0.42$, as shown in Figure
\ref{fig2} the well inside the star is  deep enough to allow one quasi-normal
mode. The number of $s$-modes increases with the depth of the well, which corresponds
to a larger stellar compactness.  
However, it should be mentioned that 
neutron stars are not expected to have such a large compactness,
unless one invokes some exotic equation of state. 
The $s$-modes are also named {\it trapped modes} because, due to the slow damping,
they are effectively trapped by the potential barrier,
and not much radiation can leak out of the star when these modes are
excited.  Axial modes on a second branch are named $w$-modes and are highly damped
(Kokkotas 1994).
The  $w$-mode frequency also depends on the stellar compactness, as we shall show in
Section \ref{wmodespolaraxial}. Therefore they carry interesting
information on the internal structure of the star. 

It should be stressed that the axial modes do not have a Newtonian counterpart.

Our approach to the polar perturbations, which couple the perturbations of the
gravitational field to those of the metric, is different from the
Newtonian approach briefly described in the introduction. Rather than focusing
on the fluid behaviour, we focus on the variables which describe 
the spacetime perturbations, assuming that, as in the case of black holes,
they are excited by the incidence of polar gravitational waves 
belonging to a particular angular harmonic.
A careful scrutiny of the structure of the polar equations
shows  that it is possible to decouple the equations
describing the metric from those describing the fluid perturbations.
This decoupling  allows us to solve the equations for the spacetime perturbations
with no reference to the motion that can be induced in the fluid, and this is
possible in general. Once the solution for the metric perturbations 
is found, the  fluid variables can be determined in terms of them by simple 
algebraic relations without further ado (Chandrasekhar \& Ferrari 1990b). 
The final set of equations to solve is described in Section 2.1.

\subsection{The equations for the polar perturbations \label{polareq}}
Assuming that the  metric which describes the unperturbed star has the form
\be
ds^{2}=e^{2\nu}(dt)^{2}-e^{2\psi}d\varphi
-e^{2\mu_{2}}(dr)^{2}-e^{2\mu_{3}}(d\theta)^{2},
\label{spa}
\ee
the functions that describe the polar
perturbations, expanded in spherical tensor harmonics and Fourier-expanded are
\beq
&\delta \nu =N_\ell(r)P_{\ell}(\cos\theta)e^{i\omega t}&
\delta\mu_2=L_\ell(r)P_{\ell}(\cos\theta )e^{i\omega t}\\
\nonumber
&\delta\mu_3=[ T_\ell(r)P_{\ell}+V_\ell(r)P_{\ell,\theta,\theta }]e^{i\omega t}&
\delta\psi =[ T_\ell(r)P_{\ell}+
V_\ell(r)P_{\ell,\theta }\cot\theta]e^{i\omega t} ,\\
\nonumber
&\delta p=\Pi_\ell(r)P_{\ell}(\cos\theta)e^{i\omega t}
&2(\epsilon +p)e^{\nu+\mu_{2}}\xi_{r}(r,\theta)e^{i\omega t}
=U_\ell(r)P_{\ell}e^{i\omega t}   \\
\nonumber
&\delta\epsilon =E_\ell(r)P_{\ell}(\cos\theta)e^{i\omega t}
&2(\epsilon
+p)e^{\nu+\mu_{3}}\xi_{\theta}(r,\theta)e^{i\omega t}
=W_\ell(r)P_{\ell,\theta}e^{i\omega t},
\eeq
where $P_\ell(\cos\theta)$ are Legendre's polynomials,  $\omega$ is the frequency,
$\delta p$ and $\delta\epsilon$ are perturbations of the pressure and of the energy
density, and 
$\xi_r,\xi_{\theta}$ are the relevant components of the Lagrangian displacement
of the generic fluid element.
Note that $(N,L,T,V)$ and $(\Pi,E,U,W)$ are, respectively, the radial part of the metric
and of the fluid perturbations. 
After separating the variables the relevant Einstein's
equations for the metric functions become
\be
\cases{
X_{\ell,r,r}+\left(\frac{2}{ r}+\nu_{,r}-\mu_{2,r}\right)
X_{\ell,r}+\frac{n}{ r^{2}}
e^{2\mu_{2}}(N_\ell+L_\ell)+\omega^{2}e^{2(\mu_{2}-\nu)}X_\ell=0,
&\cr
(r^{2}G_\ell)_{,r}=n\nu_{,r}(N_\ell-L_\ell)+
\frac{n}{ r}(e^{2\mu_{2}}-1)(N_\ell+L_\ell)
+r(\nu_{,r}-\mu_{2,r})X_{\ell,r}
+\omega^{2}e^{2(\mu_{2}-\nu)}rX_\ell,
&\cr
-\nu_{,r}N_{\ell,r}=
- G_\ell+\nu_{,r}[X_{\ell,r}+\nu_{,r}(N_\ell-L_\ell)]
+\frac{1}{ r^{2}}(e^{2\mu_{2}}-1)
(N_\ell-rX_{\ell,r}-r^{2}G_\ell)&\cr
- e^{2\mu_{2}}(\epsilon+p)N_\ell
+\frac{1}{ 2}\omega^{2}e^{2(\mu_{2}-\nu)}
\left\{ N_\ell+L_\ell+\frac{r^{2}}{ n}G_\ell+
\frac{1}{ n}[rX_{\ell ,r}+(2n+1)X_\ell ]\right\} ,&\cr
L_{\ell,r}(1-{D})+ L_\ell\left[ \left(
\frac{2}{ r}-\nu_{,r}\right)-\left(\frac{1}{ r}
+\nu_{,r}\right)D\right]+
X_{\ell,r}+X_\ell\left(\frac{1}{ r}-\nu_{,r}\right)+
{D}N_{\ell,r}+&\cr
N_\ell \left( {D}\nu_{,r}
-\frac{{D}}{ r}-{F}\right)+
\left(\frac{1}{ r}+{E}\nu_{,r}\right)\left[
N_\ell-L_\ell+\frac{r^{2}}{ n}G_\ell+
\frac{1}{ n}\left(rX_{\ell,r}+X_\ell\right)\right]=0,&\cr }
\label{compl}
\ee
where
\be
\nonumber
\cases{
{A}=\frac{1}{ 2}\omega^{2}e^{-2\nu},\qquad\quad
Q=\frac{(\epsilon +p)}{\gamma p},&\cr
\gamma=\frac{(\epsilon +p)}{p}\big (\frac{\partial p}{\partial\epsilon }
\big )_{entropy=const},\qquad\quad
{B}=\frac{e^{-2\mu_{2}}\nu_{,r}}{ 2(\epsilon+p)}
(\epsilon_{,r}-Qp_{,r}),&\cr
{D}=1-\frac{{A}}{ 2({A}+{B})}=
1-\frac{\omega^{2}e^{-2\nu}(\epsilon+p)}{
\omega^{2}e^{-2\nu}(\epsilon+p)+
e^{-2\mu_{2}}\nu_{,r}(\epsilon_{,r}-
Q p_{,r})},&\cr
{E}={D}(Q-1)-Q,&\cr
{F}=\frac{\epsilon_{,r}-Qp_{,r}}{ 2({A}+{B})}
=\frac{2\left[\epsilon_{,r}-Qp_{,r}\right](\epsilon+p)}
{2\omega^{2}e^{-2\nu}(\epsilon+p)+
e^{-2\mu_{2}}\nu_{,r}
(\epsilon_{,r}-Q p_{,r})},
&\cr}
\label{entireset}
\ee
and \op V_\ell\cl and \op T_\ell\cl have been replaced by
\op X_\ell\cl and \op G_\ell\cl  defined as
\be
\nonumber
\cases{
X_\ell=nV_\ell&\cr
\nonumber
G_\ell=\nu_{,r}[\frac{n+1}{ n}X_\ell-T_\ell]_{,r}+
\frac{1}{ r^{2}}(e^{2\mu_{2}}-1)
[n(N_\ell+T_\ell)+N_\ell]&\cr
+\frac{\nu_{,r}}{ r}(N_\ell+L_\ell)
-e^{2\mu_{2}}(\epsilon+p)N_\ell+
\frac{1}{ 2}\omega^{2}
e^{2(\mu_{2}-\nu)}[L_\ell-T_\ell+\frac{2n+1}{ n}X_\ell].&\cr}
\ee
These equations are valid in general,
also for non-barotropic equations of state.
It should be stressed that equations (\ref{compl}) govern the
variables $(X,G,N,L)$ which are  {\it metric perturbations};
however, since  the motion of
the fluid is excited by the polar perturbation, we may want to determine 
the fluid variables, ($\Pi,E,U,W$); they can be obtained in terms of the
metric functions using  the following algebraic relations
\beqn
&& W_\ell =T_\ell-V_\ell+L_\ell ,
\\
&&\Pi_\ell  = -\frac{1}{2}\omega^{2}e^{-2\nu}W_\ell-
(\epsilon +p)N_\ell ,\qquad
E_\ell  =  Q\Pi_\ell+
\frac{e^{-2\mu_{2}}}{2(\epsilon +p)}
(\epsilon_{,r}-Qp_{,r})U_\ell ,
\\
&&U_\ell =\frac{ [(\omega^{2}e^{-2\nu}W_\ell)_{,r}+
(Q+1)\nu_{,r}(\omega^{2}e^{-2\nu}W_\ell)+2(\epsilon_{,r}
-Qp_{,r})N_\ell ](\epsilon+p)}{
\left[ \omega^{2}e^{-2\nu}(\epsilon+p)+
e^{-2\mu_{2}}\nu_{,r}(\epsilon_{,r}-
Qp_{,r})\right]} .
\eeqn
Outside the star the fluid variables vanish, and the polar equations reduce
to the wave equation (\ref{reggew}) with the Zerilli potential (\ref{potzer}).

As discussed above, for the axial perturbations, the frequencies of the
quasi-normal modes 
were found by solving a problem of scattering by a central
potential;  for the polar perturbations it is not so simple,
because a Schroedinger equation holds only in the exterior of
the star, whereas a higher order system must be solved in the interior.
It is still a scattering problem, but of a more complex nature since
the incident polar gravitational waves, which excite the perturbations,
drive the fluid pulsations, which in turn emit the scattered
component of the wave.
This approach was very fruitful in many respects. First of all, given the equilibrium
configuration for any assigned equation of state,  it was very easy to
evaluate the QNM-frequency by integrating the equations for the
metric perturbations inside and outside the star, looking for the solutions which,
being regular at $r=0$, behave as pure outgoing waves at infinity.
Furthermore, we generalized the perturbed equations to slowly rotating stars, and 
derived the equations which describe how
the axial perturbations couple to the polar 
(Chandrasekhar \& Ferrari 1991b).

\subsection{Perturbed equations for a slowly rotating star \label{slowrot}}
Very briefly, the coupling mechanism is the following.
Let $Z_\ell^{0-}$ be the axial radial  function, solution of equation (\ref{axial}), 
which describes the perturbation of a non-rotating star; let 
$\epsilon(\Omega)Z_\ell^{1-}$ be the perturbation 
to first order in the star angular velocity $\Omega$.  The axial
perturbation is the sum of the two:
\[
 Z_\ell^{-}=  Z_\ell^{0-} + \epsilon(\Omega) Z_\ell^{1-}. 
\]
As $Z_\ell^{0-}$, the function $Z_\ell^{1-}$ satisfies the wave equation (\ref{axial})
with the same potential (\ref{zz1}), but with a forcing term:
\be
\sum^\infty_{\ell=2}\left\{ \frac{d^2 Z_\ell^{-1}}{dr_*^2}+
\left[\omega^2- V^-_\ell\right]  Z_\ell^{-1} \right\}
C^{-\frac{3}{2}}_{\ell+2}(\mu )
=r {e^{2\nu-2\mu_2}}(1-\mu^2)^2
\sum^\infty_{\ell=2} S_\ell^0(r,\mu ),
\ee
where $\mu=\cos\vartheta$ and $C^{-\frac{3}{2}}_{\ell+2}(\mu )$ are the Gegenbauer
polynomials.
The source term  $S_\ell^0$ is 
\[
 S_\ell^0=
\varpi_{,r}[(2  W_{\ell}^0+ N_\ell^0
+5 L_\ell^0+
2n V_\ell^0 P_{\ell,\mu}+2\mu  V_\ell^0 P_{\ell,\mu,\mu}]
+2\varpi  W_{\ell}^0(Q-1)\nu_{,r}P_{\ell,\mu};
\]
it is a combination of the functions which describe the {\it polar} perturbations on the 
{\it non-rotating} star, found by solving the equations given in Section
\ref{polareq}. It should be stressed that the coupling function $\varpi$ 
is the function responsible for the Lense-Thirring effect.
Thus a rotating star exerts a dragging not only of the bodies,
but also of the waves, and consequently an incoming polar
gravitational wave can convert, through the fluid oscillations it
excites, some of its energy into outgoing  axial waves.
This is a purely relativistic effect, and  it is due to the dragging of inertial
frames.
It is interesting to note that  the coupling between axial and polar perturbations 
satisfies rules that are similar to those known in the theory of atomic transitions:
a Laporte rule and a selection rule, according to which  the polar modes belonging to
{\it even $\ell$} can couple only with the axial modes
belonging to {\it odd $\ell$}, and conversely, and that it must be
\[
l=m+1,\qquad \hbox{or}\qquad l=m-1.
\]
Furthermore, the coupling satisfies a propensity rule (Fano 1985): the
transition  $\ell\rightarrow \ell+1$ is  strongly favoured over the
transition $\ell\rightarrow \ell-1$. 

At the time  Chandrasekhar and I wrote the series of papers on
stellar perturbations, there was a growing interest in the subject,
also motivated by the fact that
the construction of ground based interferometric detectors, 
LIGO  in the US and Virgo in Italy, had just started. 
Many studies addressed the problem of finding
the frequencies of the QNMs, to establish what kind of information they carry on the
internal structure of the emitting source. The collective effort developed using 
essentially two different perturbative approaches: one in the frequency domain, as 
for the theory developed by Thorne and collaborators or by Chandrasekhar and myself, 
another in the time domain. The time domain approach
basically consists in
separating the equations of stellar perturbations as usual in terms of spherical
harmonics, and in solving the resulting equations in terms of  two independent
variables, radial distance and time. The equations are excited 
using some numerical input, like for instance a Gaussian impulse, 
and then the QNM frequencies are found
by looking at the peaks of the Fourier transform of the signal obtained by
evolving the time-dependent equations numerically. 
A disadvantage of this evolution scheme is that one 
cannot get the complete spectrum of the QNMs either for a star or for a black 
hole. The reason is that, although any perturbation is the sum of the harmonics 
involved, in practice only a few of them can be clearly identified;
thus, to find some more modes one has to proceed empirically by changing the initial
conditions.
However, the evolution of the time dependent equations
is, still today, the only viable perturbative  method to find
the QNM frequencies and waveforms emitted by
rapidly rotating relativistic stars. To describe the problems 
which emerge when dealing with  the perturbations of a rapidly rotating star
is beyond the scope of this paper; I will discuss some related issues in the
concluding remarks.

\section{Neutron star oscillations \label{NS_QNM}}
We shall now show how the theory of perturbations of non-rotating stars 
can be applied to gain some insight into the internal structure of the emitting
source. Different classes of modes probe different aspects of the
physics of neutron stars. For instance the
fundamental mode ($f$-mode), which has been shown to be the most efficient
GW emitter by most numerical simulations,  depends on the  average density,
the pressure modes ($p$-modes) probe  the sound speed
throughout the star, the gravity modes ($g$-modes) are associated with
thermal/composition gradients and the $w$-modes are spacetime oscillations.
Furthermore, crustal modes, superfluid modes, magnetic field modes can, if present,
add to the complexity of stellar dynamics. 
The sensitivity of ground based gravitational detectors has steadily improved over
the years in a broad frequency window; the advanced version of LIGO and Virgo,
and especially third generation detectors like ET, 
promise to be  powerful instruments to detect signals emitted by oscillating stars.
The frequencies of quasi normal modes are encoded in these signals; therefore,
as the Sun oscillation frequencies are used in helioseismology to probe 
its internal structure, we hope that in the future it will be possible to 
use gravitational waves  to probe the physics of neutron stars.
One of the issues which is interesting  to address concerns  the
equation of state of matter in a neutron star core, which is actually unknown.
This problem is of particular interest, because the energies prevailing in the inner
core of a neutron star are much larger than those accessible to high 
energy experiments on Earth.
In the core, densities typically exceed  the equilibrium density of
nuclear matter, $\rho_0 =2.67 \times 10^{14}$ g/cm$^3$;
at these densities neutrons cannot be treated as non-interacting particles, and the main
contribution to pressure, which comes from neutrons, cannot be derived only from 
Pauli's exclusion principle.  Indeed,  with only this
contribution, we would find that the maximum mass of a neutron star is  $0.7~ \msun$ which,
as  observations show, is far too low.
This clearly shows that NS equilibrium requires a pressure other than the
degeneracy pressure, the origin of which  has to be traced back to the nature of
hadronic interactions. Due to the complexity of the fundamental theory of strong
interactions, the equations of state appropriate to describe a NS core
have been obtained within models, which are constrained, as much as
possible, by empirical data. They are derived within two main, different approaches:
the nonrelativistic nuclear many-body theory, NMBT, and the
relativistic mean field theory, RMFT.
In NMBT, nuclear matter is viewed as a collection of pointlike protons and
neutrons, whose dynamics is described by the nonrelativistic Hamiltonian:
\be
H = \sum_i \frac{p_i^2}{2m} + \sum_{j>i} v_{ij} + \sum_{k>j>i} V_{ijk}\ ,
\label{hamiltonian}
\ee
where $m$ and $p_i$ denote the nucleon mass
and momentum, respectively, whereas $v_{ij}$ and $V_{ijk}$ describe two- and
three-nucleon interactions. These potentials are 
obtained from fits of existing scattering data (Wiringa, Stoks \& Schiavilla 1995),
(Pudliner et al. 1995).
The ground state energy is calculated using  either variational techniques or  G-matrix
perturbation theory.
The  RMFT is based on the formalism of relativistic quantum field theory, 
nucleons are described as Dirac particles interacting through meson exchange.  
In the simplest implementation of this approach the dynamics 
is modeled in terms of a scalar and a vector field (Walecka 1974).
The equations of motion are solved in the mean field approximation, i.e. replacing the
meson fields with their vacuum expectation values, and 
the parameters of the Lagrangian density, i.e. the meson masses and coupling
constants, can be determined by fitting the empirical properties of nuclear matter, 
i.e. binding energy, equilibrium density and compressibility. 
Both NMBT and RMFT can be  generalized to take into account the appearance of 
hyperons. In the following we shall consider some EOS representative of  the two
approaches, which have been used in the literature.

It should be stressed that  
different ways of modeling  hadronic interactions affect
the pulsation properties of a star, which we are going to discuss.

\subsection{The axial and polar $w$-modes \label{wmodespolaraxial}}
\begin{figure}[ht]
\begin{center}
\includegraphics[width=13cm]{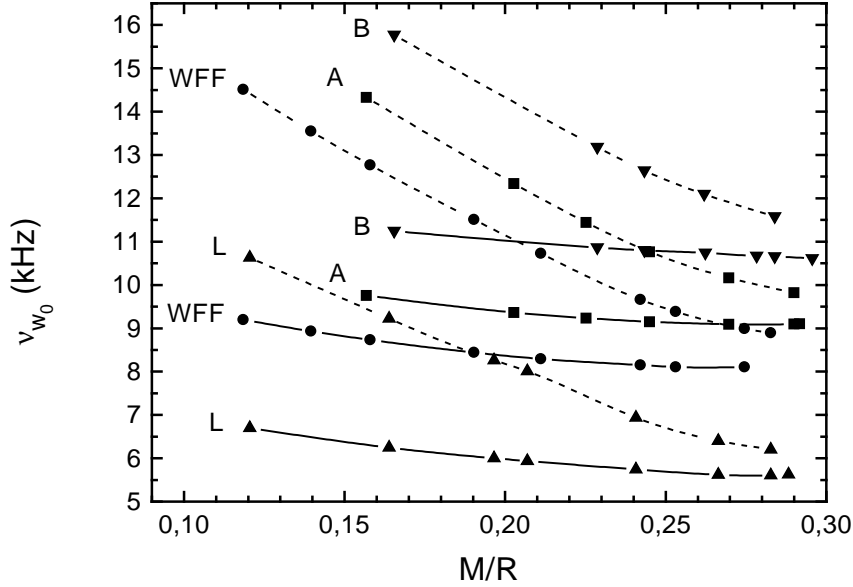}
\caption{The frequency of the first polar (dashed line) and axial
(continuous line) {w}-modes are plotted
as a function of the star compactness for the EOSs A, B, WFF, L.}
\label{fig3}
\end{center}
\end{figure}

As shown in Section \ref{secax}, the axial perturbations are described by a 
Schroedinger-like equation with a central potential barrier which depends on the
energy and pressure distribution in the unperturbed star, i.e. on
the equation of state.  
The slowly damped modes are not expected to be associated with significant gravitational
wave emission, because they are effectively trapped by the potential barrier; in
addition they appear if the star has a compactness close to the static Schwarzschild 
limit, which establishes that constant density star solutions of Einstein's 
equations exists only for $M/R < 4/9 \simeq 0.44$ .  
Conversely, the $w$-modes, which are highly damped, exist also for
stars with ordinary compactness. They have been shown to exist also for the polar
perturbations and in that case they are coupled to negligible fluid motion. 
In Figure \ref{fig3} we compare the frequencies
of the lower axial $w$-modes computed in (Benhar, Berti \& Ferrari 1999)
with those of the lower polar 
$w$-modes computed in (Andersson \& Kokkotas 1988) for several EOSs. 
The main features of different EOS are, very briefly, the following.
EOS A (Pandharipande 1971a) is pure neutron matter, with dynamics governed by a
nonrelativistic Hamiltonian containing a semi-phenomenological interaction potential.
It is obtained using NMBT. 
EOS B  (Pandharipande 1971b) is a 
generalization of EOS A, including protons, electrons and muons in 
$\beta$-equilibrium, as well as heavier baryons (hyperons and nucleon resonances) at
sufficiently high densities (NMBT). 
EOS WFF (Wiringa, Fiks \& Fabrocini 1988) is a mixture of  
neutrons, protons, electrons and muons in
$\beta$-equilibrium. The Hamiltonian includes two- and three-body interaction potentials.
The ground state energy is computed using a more sophisticated and accurate many-body
technique (NMBT). 
In EOS L (Pandharipande \& Smith 1975) neutrons interact through exchange of  mesons 
$(\omega,\rho,\sigma)$.
The exchange of heavy particles $(\omega,\rho)$ is described in terms of nonrelativistic
potentials, the effect of  $\sigma$-meson is described using relativistic field theory and
the mean-field approximation.

From Figure \ref{fig3} we see that for each selected EOS the frequency of
the {\it polar} w-modes is a rather steeply decreasing function of
the stellar compactness $ M/R$,  whereas for the {\it axial} modes
the dependence of $\nu_{w_0}$ on the compactness is weak, and
ranges within intervals that are separated for each EOS. 
This means that if an axial gravitational wave emitted by a star at a given
frequency could be detected,
we would be able to identify the equation of state prevailing in
the star's interior even without knowing its mass and radius.
Hence, the detection of axial gravitational
waves would allow us to constrain the EOS models, with
regard to both the composition of neutron star matter and the description
of the hadronic interactions.
Until very recently,  the common belief was that $w$- modes are unlikely to be  excited 
in astrophysical processes. However, it  has been shown that they are excited
in the collapse of a neutron star to a black hole,  just before the black hole forms
(Baiotti et al. 2005). 
Unfortunately the typical frequencies of these modes (of the order of several kHz)
are higher than the frequency region where the actual gravitational wave
detectors are sensitive.

\subsection{Polar Quasi Normal Modes \label{polarQNM}}
The polar metric perturbations are physically coupled to the fluid perturbations.
As shown in Section \ref{polareq},
the frequencies of the polar QNMs can be computed by solving a system of
equations involving only the metric perturbations;
however, they carry a strong imprint of the internal composition of the star,
which is present in equations (\ref{compl}) through the pressure and energy density 
profiles in the unperturbed star, which appear as coefficients of the differential
equations.
According to a   scheme  introduced by Cowling  in Newtonian gravity 
(Cowling 1942), polar modes can be classified on the basis of the restoring force which
prevails when the generic fluid element is displaced from the equilibrium
position: for  $g$-modes, or gravity modes,
the restoring force is due to buoyancy,
for  $p$-modes  it is due to pressure gradients.
The mode frequencies are  ordered as follows
\[
..\omega_{g_n} < .. < \omega_{g_1} <
\omega_f < \omega_{p_1}< .. < \omega_{p_n}..
\]
and are separated by the frequency of the fundamental mode ($f$-mode),
which has an intermediate character between $g-$ and $p-$
modes.
As  discussed in Section \ref{wmodespolaraxial},
general relativity predicts also the existence of polar $w$-modes, 
that are very weakly coupled to
fluid motion and are similar to the axial $w$-modes 
(Kokkotas \& Schutz 1992). 
The frequencies of axial and polar $w$-modes 
are typically higher than those of the fluid modes $g$, $f$ and
$p$.

\begin{figure}
\begin{center}
\includegraphics[width=7cm,angle=270]{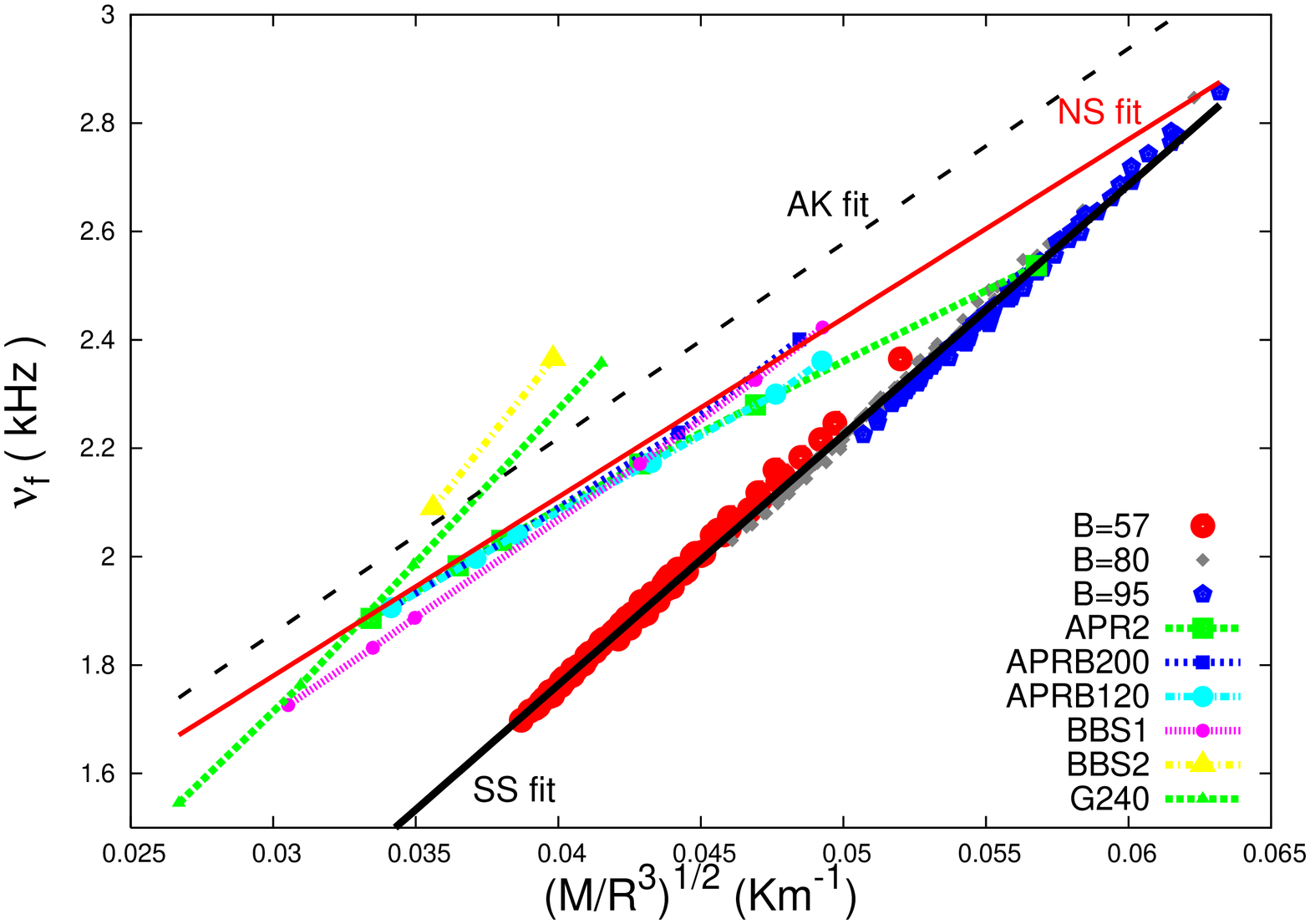}
\includegraphics[width=7cm,angle=270]{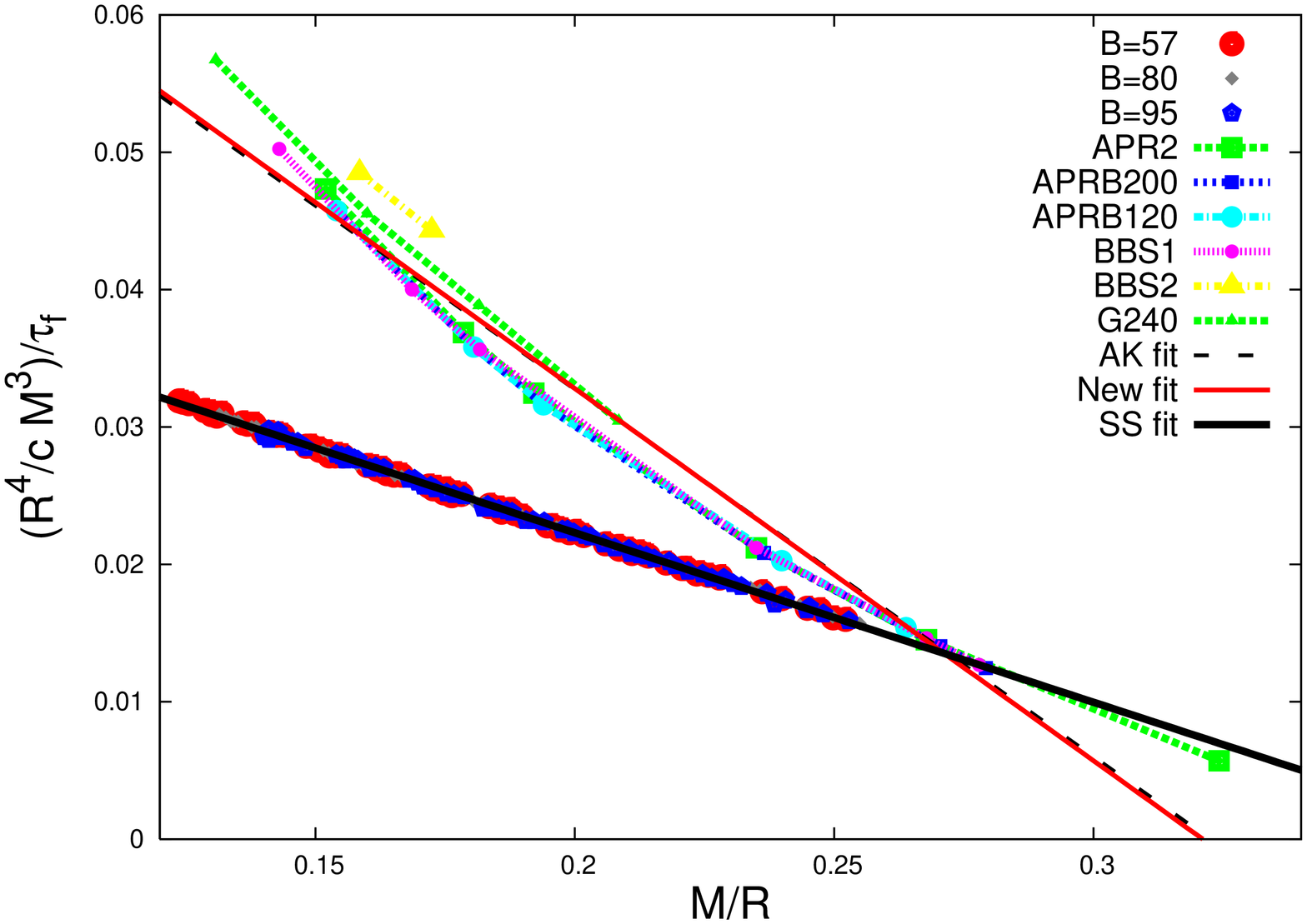}
\caption{The frequency of the fundamental mode  is plotted in the
upper panel as a function of the square root of the average density for the different
EOSs considered by Benhar, Ferrari \& Gualtieri (2004).
We also plot the fit given by Anderson \& Kokkotas  (1998) plotted as AK-fit and  our fit (NS-fit).
The NS-fit is systematically lower (about 100 Hz)
than the AK-fit. The damping time of the fundamental mode is plotted in the lower
panel as a function of the compactness $M/R$.
The AK-fit  and our fit, plotted respectively as a dashed and
continuous line, do not show significant differences.}
\label{fig4}
\end{center}
\end{figure}

If we are mainly interested in gravitational wave emission, the most interesting mode
is the $f$-mode. For mature neutron stars, its frequency is in the range  
$1-3$ kHz, which is in the bandwidth of ground based detectors Virgo and LIGO
(although not in the region where they are most
sensitive);  the damping times are of the order of a few tenths of seconds,
therefore the excitation of the $f$-mode would appear
in the Fourier transform of a gravitational wave signal
as a sharp peak and  could, in principle,
be extracted from the detector noise by an appropriate data analysis.
Moreover, the fundamental mode could be excited in several astrophysical processes,
for instance in the aftermath of a gravitational collapse, in a glitch, or due to matter
accretion  onto the star. 
For this reason, since the early years of the theory of
stellar perturbations, the interest of scientists working in this field has initially
been focussed on the determination of the  $f$-mode frequencies.
After the work of Lindblom \& Detweiler in 1983  and of Cutler \& Lindblom in 1987,
who respectively computed the $\ell=2$ and $\ell >2$
$f$-mode eigenfrequencies for the EOSs available at that time,
more recently this work has been updated, and extended to other modes,
by Anderson \& Kokkotas  (1998) and  Benhar, Ferrari \& Gualtieri (2004).
In particular, in these two papers
the $f$-mode frequency  $\nu_f$ and the corresponding damping time $\tau_f$  have been
computed to establish whether $\nu_f$ scales with the average density of the star,
as it does in
Newtonian gravity, and whether there also exists a scaling law for 
$\tau_f$. The sets of EOSs used in the two works are not identical, because the papers 
were written six years apart, although some EOSs appear in both (see the two papers for details).
The work done by Benhar, Ferrari \& Gualtieri (2004)  
 also includes examples of hybrid stars, namely
neutron stars with a core composed of quarks. 
In Anderson \& Kokkotas  (1998)  $\nu_f$ and $\tau_f$ have been fitted by a linear function of the
average density of the star $(M/R^3)^{1/2}$, and of its compactness $M/R$, as follows.
\be
\label{fitfAK}
\nu_f=0.78+1.635\sqrt{\frac{\tilde{M}}{\tilde{R}^3}},\qquad
\displaystyle\frac{1}{\tau_f}=\displaystyle\frac{\tilde{M}^3}{\tilde{R}^4}
\left[22.85-16.65\left(\frac{\tilde{M}}{\tilde{R}}\right)\right],
\ee
where $\tilde{M} = M/1.4~M_\odot$ and  $\tilde{R} = R/(10~\hbox{km})$.
Here and in the following formulae $\nu_f$  is expressed in kHz and $\tau_f$ in s.
The fits for $\nu_f$ and $\tau_f$ 
obtained by Benhar, Ferrari \& Gualtieri (2004) using the new set of EOSs are
\be
\label{fitf}
\nu_f=a+b\sqrt{\frac{M}{R^3}},\quad
a=0.79\pm0.09~ (\hbox{in~kHz}),\quad b=33\pm2~ (\hbox{in~ km}) ,
\ee
and
\beq
\label{fitauf}
\displaystyle\frac{1}{\tau_f}=\displaystyle\frac{c M^3}{R^4}
\left[a+b\left(\frac{M}{R}\right)\right],\qquad
a=[8.7\pm0.2]\cdot 10^{-2},\qquad
b=-0.271\pm0.009\,.
\eeq
In equations (\ref{fitf}) and (\ref{fitauf}) mass and radius are in km
(i.e. mass is multiplied by G/$c^2$) and $c=3\cdot10^5$ km/s.
The data for the different EOSs used by Benhar, Ferrari \& Gualtieri (2004),
and the fits given in equations (\ref{fitfAK})-(\ref{fitauf})
are shown in Figure \ref{fig4}.
$\nu_f$  is plotted in the upper panel as a function of the average density;
the fit (\ref{fitfAK}) is shown as a black dashed line labelled `AK fit',
whereas the new fit is indicated as a red continuous line labelled `NS fit'.
The NS fit is lower by about 100 Hz than the AK fit, showing that the new
EOSs are, on average, less compressible than the old ones.
The quantity $(R^4/cM^3)/\tau_f$ given in equation (\ref{fitauf})
is plotted in the lower panel 
of Figure \ref{fig4} versus the stellar compactness $M/R$.
In this case the AK fit for $\tau_f$ (\ref{fitfAK}),
 and  the NS fit (\ref{fitauf}) are nearly coincident.
For comparison, in both panels of Figure \ref{fig4} we show the frequency and the
damping time of the $f$-mode of a population of strange stars, namely stars entirely
made of up, down and strange quarks, modeled using the MIT
Bag-model, spanning the allowed range of
parameters, which are the Bag constant, the coupling constant $\alpha_S$ and
the quark masses (see Benhar et al. 2007 for details).
The parameters of the fits for strange stars are
\be
\label{fitf_strange}
\hbox{for } ~\nu_f\qquad
a=-[0.8\pm0.08]\cdot 10^{-2}~,\quad b=46\pm0.2 ,
\ee
and
\be
\label{fitauf_strange}
\hbox{for } ~\tau_f\qquad
a=[4.7\pm 5\cdot 10^{-3}]\cdot 10^{-2},\qquad
b=-0.12\pm 3\cdot 10^{-4}\,.
\ee
In Figure \ref{fig4} the fits for strange stars are labelled
as `SS fit'. It is interesting to note
that the SS fits are quite different from those appropriate
for neutron stars (AK- and NS-fits). First of all the errors on the parameters are much
smaller, which indicate that the linear behaviour is followed by these stars,
both for $\nu_f$ and for $\tau_f$, 
irrespective of the values of the parameters of the model. Moreover, the difference
between the fits is much larger for lower values of the average density.

The empirical relations given in eqs. (\ref{fitfAK})-(\ref{fitauf_strange})
could be used to constrain the values of the star mass and radius, were
the values of  $\nu_f$ and $\tau_f$ identified in a detected gravitational
signal. The stellar parameters would be further constrained if other modes are
excited and detected and, knowing them, we would gain information on 
the equations of state of
matter in the neutron star core, whose uncertainty is due, as explained earlier, to
our ignorance of hadronic interactions.
Furthermore, if the neutron star mass is known, as it may be 
if the star is in a binary system,
the detection of a signal emitted by the star oscillating in the $f$-mode may provide
some further interesting information (Benhar et al. 2007). 
\begin{figure}
\begin{center}
\includegraphics[width=7cm,angle=270]{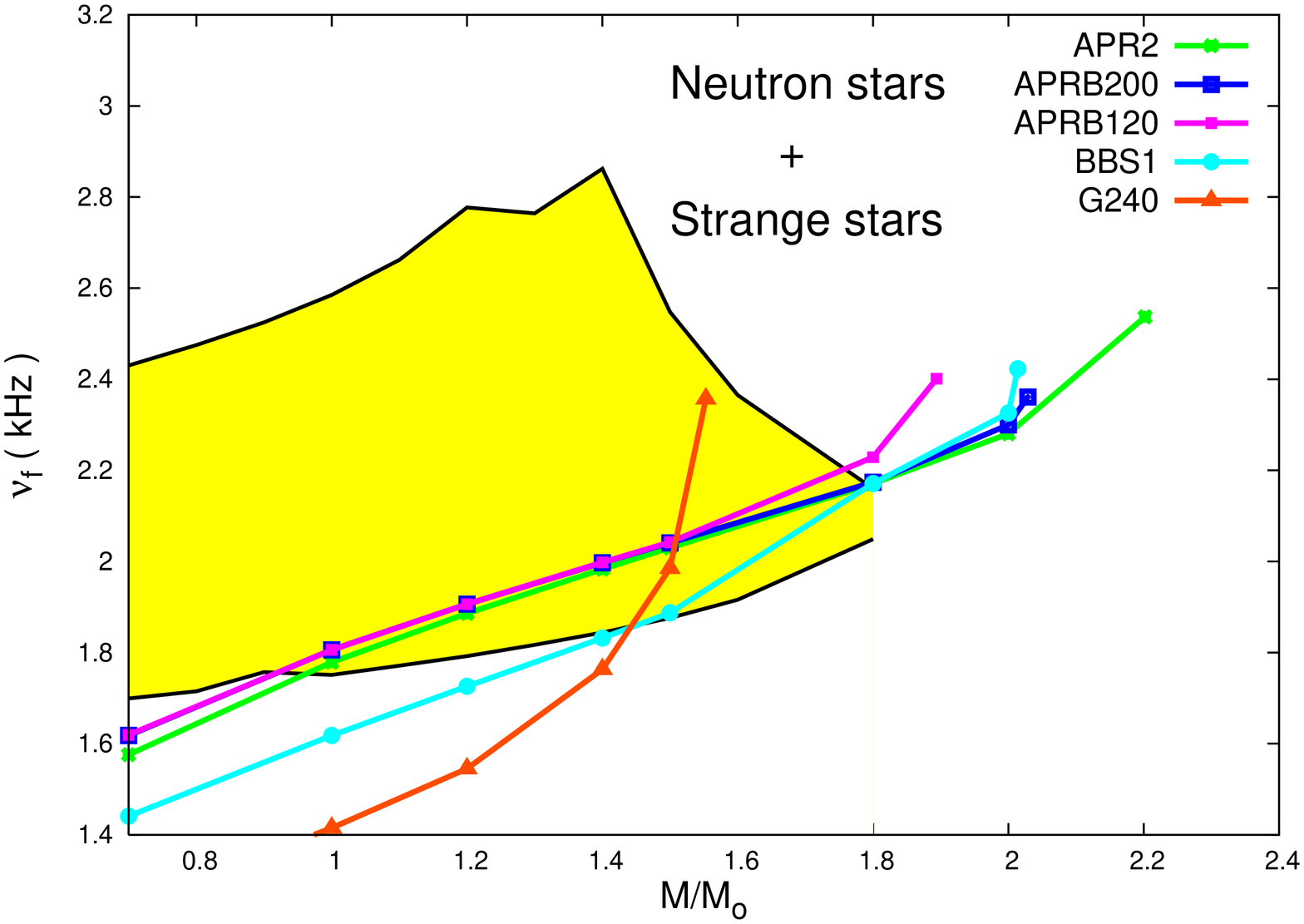}
\caption{The frequency of the fundamental mode is
plotted as a function of the mass of the star, for neutron/hybrid stars
(continuous lines)
and for strange stars modeled using the MIT bag model, spanning the set of
parameters indicated in the range allowed by high energy experiments (dashed region).}
\label{fig5}
\end{center}
\end{figure}
In Figure \ref{fig5} we plot $\nu_f$ as a function of the stellar mass,
for neutron/hybrid stars and for strange stars modeled using the MIT bag model.
Note that $1.8~M_\odot$ is the maximum mass above which no stable strange star can exist.
We see that there is a small range of frequency where neutron/hybrid stars are
indistinguishable from strange stars; conversely,  there is a large
frequency region where only strange stars can emit.
Moreover, strange stars cannot emit gravitational waves
with $\nu_f \lesssim 1.7$ kHz, for any value of the mass in the range
we consider.
For instance, if the stellar mass is  $M=1.4~ M_\odot$, 
a signal with $\nu_f \gtrsim 2$~kHz would belong to a strange star.
Figure \ref{fig5} also shows that, even if we do not know the mass of the star
(as it is often the case for isolated pulsars), if
$\nu_f \gtrsim 2.2$ kHz, apart from a very narrow region of masses where
stars with hyperons would emit (EOS BBS1 and G240),
we can reasonably rule out that the signal is emitted by a neutron star.
In addition, it is possible to show that, since $\nu_f$ is an increasing 
function of the Bag constant $B$, if a signal emitted by an oscillating
strange star were detected,
it would be possible to set constraints on $B$ much more stringent than those
provided by the available experimental data (Benhar et al. 2007).

In conclusion, the QNM frequencies can be used  to gain
direct information on the equation of state of matter in a neutron star core.

The crucial question now is: do we have a chance to detect a signal emitted by
a star oscillating in a polar quasi-normal mode?
Detection chances depend on
how much energy is  channeled into the pulsating  mode, which is unknown,
and on whether the mode frequency is in the detector bandwidth.
The signal emitted by a star pulsating in a given mode of frequency $\nu$ and damping
time $\tau$, has the form of a damped sinusoid
\begin{equation}
h(t) = {\cal A} e^{-(t-t_0)/\tau} \sin [ 2\pi \nu (t-t_0)] \quad \mbox{ for }
       t >  t_0,
\end{equation}
where $t_0$ is the arrival time of the signal at the detector (and
$h(t)=0$ for $t<t_0$).
The wave amplitude ${\cal A}$  can be expressed in terms of the
energy radiated in the oscillations,
\begin{equation}
{\cal A} \approx 7.6\times 10^{-24} \sqrt{{\Delta E_\odot \over
10^{-12} } {1 \mbox{ s} \over \tau}}
 \left( {1 \mbox{ kpc} \over d } \right)
\left({ 1 \mbox{ kHz} \over \nu} \right)  \ .
\end{equation}
where $\Delta E_\odot = \Delta E_{GW}/M_\odot c^2$.
This quantity is unknown. Therefore to assess the detectability of a signal we can
only evaluate how much energy should be emitted in a given mode, in order for the signal 
to be detected by a given detector with an assigned signal to noise ratio $(S/N)$ 
\begin{equation}
\left({S \over N} \right)^2 = { 4Q^2 \over 1+4Q^2} {{ \cal A}^2 \tau
\over 2S_n}.
\label{sign}\end{equation}
In this equation $Q=\pi \nu \tau$ is the quality factor and
$S_n$ is the detector spectral noise density. 
Assuming  as a bench-mark for $\Delta E_\odot$ the energy involved in a typical
pulsar glitch, in which case  a mature neutron star might  radiate
an energy of the order of ${\Delta E_{GW}} = 10^{-13}M_\odot c^2$,
and assuming ${\nu}\sim 1500$ Hz, $\tau\sim 0.1$ s, $d=1$ kpc, we find
 ${\cal A} \approx 5\times 10^{-24}$.
Such a signal is too weak to be seen by actual detectors, therefore we conclude that
3rd generation detectors are needed to detect signals from old neutron
stars.  More promising are the oscillations of newly-born neutron stars;
indeed, since a NS forms as a consequence of  a violent, and generally
non-symmetric event -- the gravitational collapse -- a fraction of its large
mechanical energy may go into non-radial oscillations and would be radiated 
in gravitational waves. Thus, during the first few seconds of the NS life 
more energy could be stored in the pulsation modes than when the star is cold and
old.
In addition, during this time the star is less dense than at the end of the
evolution;  consequently, the frequencies of the modes which depend on the
stellar compactness (as for instance the $f$-mode) are lower and therefore
span a frequency range were the detectors are more
sensitive (Ferrari, Miniutti \& Pons 2003).
For instance, if we assume that an energy 
$\Delta E_{GW} = 1.6 \cdot 10^{-9}~M_\odot c^2$
is stored in the $f$-mode of a neutron star just formed in the Galaxy, 
the emitted signal would be detectable with a signal to noise ratio 
 ${S/N= 8}$ by advanced Virgo/LIGO, and with 
 ${S/N= 2.7}$ by Virgo+/LIGO, the upgraded configurations now in operation.

\section{Stellar perturbations and magnetar oscillations}
Magnetars are neutron stars whose magnetic field is, according to current models,
as large as $10^{15}$ G (Thompson \& Duncan 1993, 2001). During the last 
three decades some  very interesting astrophysical
events have been observed which are connected to magnetar activity and stellar
pulsations. They involve Soft Gamma Repeaters (SGRs), which are thought to be
magnetars; these sources occasionally release bursts of huge amount of energy 
($L \simeq 10^{44}-10^{46}$ ergs/s), and 
these giant flares are thought of being 
generated from large-scale rearrangements of the inner field, or
catastrophic instabilities in the magnetosphere (Thompson \& Duncan 2001; 
Lyutikov 2003).
Up to now, three of these events have been detected: 
SGR 05026-66 in 1979, SGR 1900+14 in 1998
and SGR 1806-20 in 2004. In two of them  (SGR 1900+14 and SGR
1806-20), a tail lasting several hundred seconds has been observed, and 
a detailed study of this part of the spectrum has revealed the presence of 
quasi-periodic oscillations (QPOs) with frequencies 
\[
18,~ 26, ~30, ~92, ~150, ~625\quad\hbox{and}\quad 1840 \quad\hbox{Hz}\quad
\]
for SGR 1806-20 (Watts \& Strohmayer 2006), and
\[
28,~ 53,~ 84 \quad\hbox{and}\quad 155 \quad\hbox{Hz}\quad\]
for SGR 1900+14 (Strohmayer \& Watts 2006).
The discovery of these oscillations stimulated an interesting and lively debate
(still ongoing) among groups working on stellar perturbations, about the 
physical origin of these sequences.
Of course in order to study these oscillations the magnetic field and its 
dynamics have to be included in the picture, 
while rotation plays a less important role, since observed magnetars 
are all very slowly rotating.  
The problem presents extreme complexity both at conceptual and at  computational
levels; therefore it is usually approached using simplifying assumptions
and/or approximations.
Some of the studies try to explain the
observed modes in terms of torsional oscillations of the crust
(Samuelsson \& Andersson 2007; Sotani, Kokkotas \& Stergioulas 2007), others
attribute  the observed spectra to global magneto-elastic oscillations 
(Glampedakis, Samuelsson \& Andersson 2006), still
others investigate the interaction between the torsional
oscillations of the magnetar crust and a continuum of 
magnetohydrodynamic modes (the Alfven continuum) in the fluid core
(Levin 2007; Sotani, Kokkotas \& Stergioulas 2008;
Colaiuda, Beyer \& Kokkotas 2009; Cerd\'a-Dur\'an, Stergioulas \& Font
2009; Colaiuda \& Kokkotas 2010; Gabler et al. 2011)
using different approaches and approximations.
In particular, in (Colaiuda \& Kokkotas 2010) the torsional oscillations of a
magnetar have been studied in a general relativistic framework, perturbing
Einstein's equations  in the Cowling
approximation, i.e. neglecting gravitational field perturbations.
By this approach  the crust-core coupling due to the strong magnetic field 
has been shown to be able to explain the origin of the observed frequencies,
at least for SGR 1806-20, if a suitable stellar model is considered.
With this identification, constraints on the mass and radius of the star, and
consequently on the EOS in the core, can be set; estimates of
the crust thickness and of the value of
the magnetic field at the pole can also be inferred.

Thus,  the theory of stellar perturbations has been 
generalized  to magnetized stars, although for now
only with considerable restriction, since
only torsional oscillations have been considered (i.e. axial perturbations)
and only in the Cowling approximation. Nevertheless, it already provides very interesting 
information on the dynamics of these stars and allows us to confront 
the predictions  with  astronomical observations. 

\section{Concluding remarks}
I would like to conclude this review by mentioning the fact that 
the theory of perturbations of rotating stars has not been developed 
to the same extent as the theory of non-rotating stars. 
The main reason is that the mathematical tools appropriate for a successful variable
separation has not been found yet.
When the perturbations of a non-rotating black hole are studied, separation of
variables is achieved by expanding all tensors in tensorial spherical harmonics.
In the case of Kerr perturbations, namely of perturbations of an axisymmetric,
Petrov type D background, the same result is obtained by expanding the Newman-Penrose 
quantities in  oblate spheroidal harmonics.
When perturbing a rotating star, i.e.  an axisymmetric solution of Einstein+hydro
equations, an expansion in terms of tensorial spherical harmonics leads, 
as we have seen in the case of slow rotation in Section \ref{slowrot},
 to a coupling between polar and axial perturbations.
If rotation is not slow, the number of couplings to be considered increases 
to such an extent that the problem becomes untreatable, both from a
theoretical and from a computational point of view.
One may argue that, since the background of a rapidly rotating star is 
not spherically symmetric, tensorial spherical harmonics are unappropriate, and
this is certainly true. However, even if we try to use the oblate spheroidal
harmonics and the Newmann Penrose formalism  we fail: the coupling between 
the metric and the fluid makes the separation impossible, at least in terms 
of these harmonics (unlike the Kerr metric,
the metric describing a star is not of Petrov type D).
For this reason, perturbations of rotating stars have been studied either
in the slow rotation regime, or using the Cowling
approximation, which neglects  spacetime perturbations, or using other simplifying
assumptions. For instance,
as far as the mode calculation is concerned, the Cowling approximation allows 
determination with reasonable accuracy
the frequency of the higher order $p$-modes, of the $g$-modes and of the
inertial modes, like the $r$-modes, thus allowing us to gain information on the
onset of related instabilities.
Conversely, the determination of the $f$-mode frequency, 
which is so important from the point of
view of gravitational wave emission, is not very precise,
leading to errors as large as $\sim$20\%.

However, it should be mentioned that non-linear simulations of rotating stars
are producing very interesting results; for instance in a recent paper
Stergioulas and collaborators (Zink et al. 2010)
have been able to follow the frequency of the
non-axisymmetric fundamental mode of a sequence of rotating stars with increasing 
angular velocity, up to the onset of the CFS instability, making also very
optimistic estimates of the amount of gravitational radiation which 
could be emitted in the process.
To describe matter in the neutron star they use a 
simple model (a polytropic equation of state and 
uniform rotation); however, their result  indicates
 that numerical relativity is making giant steps in this field.
Thus, supercomputers are making  accessible very complex problems, which 
only ten years ago one would not have dreamed of solving; however,
perturbation theory still remains a very powerful tool to investigate many physical 
problems and it should be used in parallel with the numerical work
to gain a deeper insight into the physics of stellar oscillations.



\label{lastpage}

\begin{thebibliography}{99}
%
\bibitem{andresson1994} Andersson N.,  1994, CQG, 11, 3003
\bibitem{Andersson_Araujo_Schutza} Andersson N., Araujo M.E., Schutz B.F.,
1993a, CQG, 10, 735
\bibitem{Andersson_Araujo_Schutzb} Andersson N., Araujo M.E., Schutz B.F.,
1993b, CQG, 10, 757
\bibitem{Andersson_Araujo_Schutzc} Andersson N., Araujo M.E., Schutz B.F.,
1993c, Phys. Rev. D, 49, 2703
\bibitem{andkok1998} Andersson N., Kokkotas K.D., 1998,  MNRAS, 299, 1059

\bibitem{Andersson_reggepoles} Andersson N., Thylwe K., 1994, CQG,
11, 2991;
\bibitem{BHRS} Baiotti L., Hawke I., Rezzolla L., Schnetter E., 2005,
 Phys. Rev. Lett., 94, 131101
\bibitem{omarema} Benhar O., Berti E., Ferrari V., 1999, MNRAS,  310, 797
\bibitem{astero_nostro} Benhar O., Ferrari V., Gualtieri L., 2004,
 Phys. Rev. D70, 124015
\bibitem{strangestars}
Benhar O., Ferrari V.,  Gualtieri L.,  Marassi S., 2007, GRG, 39, 1323

\bibitem{Campolattaro_Thorne}Campolattaro A., Thorne K.S., 1970,  ApJ, 159, 847
\bibitem{Cerda-Duranetal2009} Cerd\'a-Dur\'an P., Stergioulas N., Font J. A., 2009,
MNRAS, 397, 1607

\bibitem{MT} Chandrasekhar S., 1984, The mathematical theory of black holes,
Claredon Press, Oxford
\bibitem{chadet} Chandrasekhar S., Detweiler S.L., 1975,  Proc. R. Soc.  Lond., A344, 441
\bibitem{chf0} Chandrasekhar S.,  Ferrari V., 1990a, Proc. R. Soc.  Lond., A428, 441
\bibitem{chf2}
Chandrasekhar S.,  Ferrari V., 1990b, Proc. R. Soc.  Lond., A432, 247
\bibitem{chf1} Chandrasekhar S.,  Ferrari V., 1991a, Proc. R. Soc.  Lond., A435, 645
\bibitem{chf3}
Chandrasekhar S.,  Ferrari V., 1991b, Proc. R. Soc.  Lond., A433, 423
\bibitem{chf4}
Chandrasekhar S.,  Ferrari V., 1991c, Proc. R. Soc.  Lond., A434, 449
\bibitem{chf6}
Chandrasekhar S.,  Ferrari V., 1992, Proc. R. Soc.  Lond., A437, 133
\bibitem{chf5}
Chandrasekhar S.,  Ferrari V., Winston R., 1991, Proc. R. Soc.  Lond., A434, 635

\bibitem{Colaiudaetal2009} Colaiuda A., Beyer H., Kokkotas K.D., 2009, MNRAS, 396,
1441
\bibitem{Colaiuda_kokkotas2010} Colaiuda A., Kokkotas K.D., 2010, arXiv:1012.3103v2
\bibitem{cowling} Cowling T.G., 1942, MNRAS, 101, 367

\bibitem{Cutler_Lindblom_1987} Cutler C., Lindblom L., 1987, ApJ, 314, 234
\bibitem{Fano1985} Fano U., 1985, Phys. Rev. A, 32, 617

\bibitem{ferrari_miniutti_pons_2003}
Ferrari V., Miniutti G., Pons J.A.,  2003,   MNRAS,  342, 629
\bibitem{ferrari_mashoona} Ferrari V., Mashhoon B., 1984a, Phys. Rev. D, 30, 295
\bibitem{ferrari_mashoonb} Ferrari V., Mashhoon B., 1984b, Phys. Rev.Lett, 52, 1361
\bibitem{gableretal2011}
Gabler M., Cerd\'a Dur\'an P., Font J.A., Muller E., Stergioulas N., 2011, MNRAS,
410, L37
\bibitem{glampe_etal_2006}
Glampedakis K., Samuelsson L., Andersson N., 2006, MNRAS, 371, L74
\bibitem{Ipser_Thorne} Ipser J.R., Thorne K.S., 1973,  ApJ, 181, 181
\bibitem{ll} Landau L.D., Lifshitz E.M., 1975, The classical theory of fields, 
New York: Pergamon Press
\bibitem{levin_2007} Levin Y., 2007, MNRAS, 377, 159

\bibitem{Lindblom_detweiler_1983}Lindblom L.,  Detweiler S.L., 1983, ApJ Suppl., 53, 73
\bibitem{Lyutikov} Lyutikov M., 2003, MNRAS, 346, 540
\bibitem{wkostas} Kokkotas K.D., 1994,  MNRAS,  268, 1015
\bibitem{kokkoschutzwmodes} Kokkotas K.D., Schutz B.F., 1992, MNRAS,  255, 119

\bibitem{Pandaripande_1971a} Pandharipande V.R., 1971a, Nucl. Phys. A, 174, 641
\bibitem{Pandaripande_1971b} Pandharipande V.R., 1971b, Nucl. Phys. A, 178, 123
\bibitem{PandaripandeSmith1975} Pandharipande V.R., Smith R.A., 1975, Phys. Lett., 59, 15

\bibitem{pr1} Press W.H., 1971,  ApJ, 170, L105
\bibitem{PPCPW} Pudliner B.S., Pandharipande V.R., Carlson J., Pieper S.C.,
Wiringa R.B., 1995, Phys. Rev. C, 56, 1720

\bibitem{regwe} Regge T.,  Wheeler J.A., 1957, Phys. Rev.,   108, 1063
\bibitem{andersson_Samuelsson} Samuelsson L., Andersson N., 2007, MNRAS, 374, 256
\bibitem{schutz_will} Schutz B.F., Will C.M., 1985, ApJ, 291, L33-L36
\bibitem{sorkinflux} Sorkin R., 1991, Proc. R. Soc. Lond., A435, 635

\bibitem{Sotani_kokkotas_stergioulas_2007}
Sotani H., Kokkotas K.D., Stergioulas N., 2007, MNRAS, 375, 261
\bibitem{Sotanietal2008} Sotani H., Kokkotas K.D., Stergioulas N., 2008, MNRAS,
385,L5

\bibitem{Watts_Strohmayer2}  Strohmayer T.E., Watts A.L., 2006, ApJ 653,  593
\bibitem{teukolskya} Teukolsky S., 1972,  Phys. Rev. Lett.,  29, 1114
\bibitem{teukolskyb} Teukolsky S., 1973, ApJ,  185, 635
\bibitem{Thompson_Duncan1993}Thompson C., Duncan R.C., 1993, ApJ, 408, 194 
\bibitem{Thompson_Duncan2001}Thompson C., Duncan R.C., 2001, ApJ, 561, 980
\bibitem{Thorne_1} Thorne K.S., 1969a,  ApJ,  158, 1
\bibitem{Thorne_2} Thorne K.S., 1969b,  ApJ, 158, 997
\bibitem{Campolattaro_Thorne_1} Thorne K.S., Campolattaro A., 1967, ApJ, 149, 591
\bibitem{Campolattaro_Thorne_2} Thorne K.S., Campolattaro A., 1968, ApJ, 152, 673
\bibitem{vish} Vishveshwara C.V., 1970, Phys. Rev.D, 1, 2870
\bibitem{QHD1} Walecka J.D., 1974, Ann. Phys., 83, 491
\bibitem{Watts_Strohmayer1} Watts A.L, Strohmayer  T.E., 2006, ApJ, 637,  L117
\bibitem{WiringaFiksFabrocini_1988} Wiringa R.B., Fiks V., Fabrocini A., 1988,
Phys. Rev. C, 32, 1057
\bibitem{WSS} Wiringa R.B., Stoks V.G.J.,  Schiavilla R., 1995, Phys. Rev. C,  51, 38

\bibitem{zerillia} Zerilli J.F., 1970a,  Phys. Rev. D,  2, 2141;
\bibitem{zerillib} Zerilli J.F., 1970b,  Phys. Rev. Lett.,  24, 737
\bibitem{stergioulasetal2010} Zink B., Korobkin O., Schnetter E.,  Stergioulas N.,
2010, Phys. Rev. D81, 084055


\end{thebibliography}
\end{document}